\documentclass[12pt]{article}
\usepackage[dvips]{graphicx}
\usepackage{pdproc}

  \makeatletter 
  \def\@cite#1{[#1]} 
  \makeatother    
\begin{document}

%%%%%%%%%%%%%%%%%%%%%%%%%%%
\newcommand{\beeq}{\begin{equation}}
\newcommand{\eneq}{\end{equation}}
\newcommand{\beqn}{\begin{eqnarray}}
\newcommand{\eeqn}{\end{eqnarray}}

\def\mybig{\displaystyle \strut }
\def\mbig{\displaystyle }
\def\dd{\partial}
\def\la{\raise.16ex\hbox{$\langle$}\lower.16ex\hbox{}  }
\def\ra{\, \raise.16ex\hbox{$\rangle$}\lower.16ex\hbox{} }
\def\go{\rightarrow}
\def\yield{\Longrightarrow}
\def\next{{~~~,~~~}}
\def\onehalf{ \hbox{${1\over 2}$} }
\def\onethird{ \hbox{${1\over 3}$} }

\def\GUT{{\rm GUT}}
\def\eff{{\rm eff}}

\def\myfrac#1#2{{\mybig #1\over \mybig #2}}
\def\myitem#1#2{\hbox{\hskip .4cm \vtop{\hsize .5cm #1}\vtop{\hsize 15.5cm  #2}}\vskip .1cm }
\def\ignore#1{}

%%%%%%%%%%%%%%%%%%%%%%%%%%%%

\renewcommand{\thefootnote}{\alph{footnote}}

\title{
Dynamical Gauge-Higgs Unification (SUSY04)\footnote{To appear in the Proceedings 
of {\it ``SUSY 2004''}, Tsukuba, Japan, June 17-23, 2004}
}

\author{YUTAKA HOSOTANI}

\address{ 
Department of Physics, Osaka University\\
Toyonaka, Osaka 560-0043, Japan
%%%%% You may comment out the e-mail address line below.  
\\ {\rm E-mail: hosotani@phys.sci.osaka-u.ac.jp}}

\abstract{
Dynamical gauge-Higgs unification is presented in higher dimensional
gauge theory, in which both adjoint and fundamental Higgs fields 
are a part of gauge fields. Dynamical gauge symmetry breaking is 
induced through the Hosotani mechanism.  Gauge theory, including 
the $U(3) \times U(3)$ model, is examined on $M^4 \times (T^2/Z_2)$.
 \hfill (OU-HET 479/2004)}

\normalsize\baselineskip=15pt

\section{Gauge-Higgs unification}

Gauge theory in higher dimensions has been studied extensively
in which Higgs bosons in four dimensions can be identified with 
extra-dimensional components of gauge fields.
The idea of unifying Higgs scalar fields with gauge fields was first 
put forward by Manton.\cite{Manton1} 
In the $SU(3)$ or $G_2$ gauge theory on $M^4 \times S^2$,
the gauge symmetry breaks down to
the electroweak $SU(2)_L \times U(1)_Y$ by 
nonvanishing field strengths  on $S^2$, which at the same time 
induce the instability due to the nonvanishing energy density.

A better scenario is to start with gauge theory on non-simply
 connected space.  It was shown that dynamics of Wilson line phases,
 which at the classical level give vanishing energies,  can 
  induce gauge symmetry breaking at the quantum level.\cite{YH1,YH2}
Adjoint Higgs fields in grand unified theories (GUT) are identified 
with extra-dimensional 
components of gauge fields. Dynamical symmetry breaking such as 
$SU(5) \go SU(3) \times SU(2) \times U(1)$ takes place at the quantum level.

Recently the scenario has been elaborated on orbifolds in which
boundary conditions  play  an additional role.  Chiral fermions
are incorporated and the gauge hierarchy problem in GUT is 
solved.\cite{orbifold1,Antoniadis,HHHK}
Different sets of boundary conditions, however,  
can be equivalent through the Hosotani mechanism.  
Quantum treatment of Wilson line phases becomes crucial to 
determine the physical symmetry of the theory.

There are two types of gauge-Higgs unification.~\cite{YH4}

\noindent (i) $\underline{\hbox{Gauge-adjoint-Higgs unification}}$
%\subsection{Gauge-adjoint-Higgs unification}

%\myitem{}{
In  GUT, Higgs fields in the adjoint 
representation are responsible for reducing gauge symmetry to the 
standard model symmetry, $SU(3) \times SU(2) \times U(1)$.   
In higher dimensional gauge theory extra-dimensional components of 
gauge fields serve as Higgs fields in the adjoint representation 
in four dimensions at low energies. This is the
gauge-adjoint-Higgs unification introduced in ref.\ \cite{YH1}. 
%}

\noindent (ii) 
$\underline{\hbox{Gauge-fundamental-Higgs unification}}$

%\myitem{}{
Electroweak symmetry breaking is induced by Higgs fields in the 
fundamental representation.  They  have another important role of 
giving fermions finite masses. 
To unify a scalar field in the fundamental representation 
with gauge fields, the gauge group has to be enlarged.
In Manton's approach,\cite{Manton1}  the gauge group is $SU(3)$ 
or $G_2$. In GUT one can start with $SU(6)$ which breaks to $SU(3) \times SU(2) \times  U(1)^2$.~\cite{gaugeHiggs3}
%}

\section{Gauge theory on  orbifolds}

If the space is non-simply connected, Wilson line phases become
physical degrees of freedom.  They are dynamical and affect physics.
At the classical level Wilson line phases label degenerate vacua.
The degeneracy is lifted by quantum effects. 
If the effective potential of Wilson line phases is 
minimized at nontrivial values of Wilson line phases, then the 
rearrangement of gauge symmetry takes place. 

Consider $SU(N)$ guage theory on $M^4 \times (T^2/Z_2)$.  
Let $x^\mu$  and $\vec y = (y_1, y_2)$  be
coordinates of $M^4$ and $T^2$, respectively.
$\vec y$ and $\vec y + \vec l_a$ ($a=1,2$) are identified
on $T^2$ where $\vec l_1 = (2\pi R_1, 0)$ and 
$\vec l_2 = (0, 2\pi R_2)$.  The $Z_2$-orbifolding is obtained 
by identifying $-\vec y$ with $\vec y$. There appear four
fixed points on $T^2/Z_2$ under the parity $\vec y \go -\vec y$ ;
$\vec z_0=\vec 0$, $\vec z_1=\onehalf \vec l_1$, 
$\vec z_2=\onehalf \vec l_2$,  and 
$\vec z_3=\onehalf (\vec l_1+ \vec l_2)$.

Gauge fields satisfy the following boundary conditions.~\cite{HNT1}
\beqn
&&\hskip -1cm 
A_M(x, \vec y + \vec l_a) =
U_a A_M(x, \vec y ) \, U_a^\dagger  \quad
  (a=1,2) ,\cr
\noalign{\kern 10pt}
&&\hskip -1cm 
\pmatrix{A_\mu \cr A_{y_a} \cr} (x, \vec z_j - \vec y) =
P_j \pmatrix{A_\mu \cr - A_{y_a} \cr} (x, \vec z_j + \vec y) \, 
   P_j^\dagger \quad (j=0,1,2,3) \cr
\noalign{\kern 10pt}
&&\hskip -1.cm 
U_a, P_j  \in SU(N) ~~,~~
[U_1 , U_2 ] = 0 ~~, ~~
P_j^\dagger = P_j = P_j^{-1}  ~.
\label{BC1}
\eeqn
In gauge theory $U_a$ and $P_j$ need not be the identity matrix.
The only requirement is that physical quantities, particularly
the Lagrangian density, should be single-valued on $T^2/Z_2$.  
Not all of $U_a$ and $P_j$ are independent;
\beeq
U_a = P_a P_0 ~~,~~ P_3 = P_1 P_0 P_2 = P_2 P_0 P_1 ~~.
\label{BC2}
\eneq
Similarly fermion fields satisfy
\beqn
&&\hskip -1cm 
\psi(x, \vec y + \vec l_a) 
  =  \eta_0 \eta_a \, T[U_a] \psi(x, \vec y) ~~~, \cr
\noalign{\kern 10pt}
&&\hskip -1cm 
\psi(x, \vec z_j - \vec y) = \eta_j \, T[P_j] \, (i\Gamma^4\Gamma^5)
\psi(x, \vec z_j + \vec y) \qquad (\eta_j = \pm 1) ~.
\label{BC3}
\eeqn
$T[U_a] \psi= U_a \psi$ or $U_a \psi U_a^\dagger$ for $\psi$ 
in the fundamental or adjoint representation, respectively.

Different sets of boundary conditions can be gauge equivalent.
Under a gauge transformation
$A_M' =\Omega A_M \Omega^{\dagger} 
-(i/ g)\Omega\partial_M \Omega^{\dagger}$,
$A_M'$ obeys a new set of boundary conditions $\{ P_j' , U_a' \}$ where
\beqn
&&\hskip -1cm
P_j' =\Omega(x, \vec z_j -\vec y) \, P_j 
   \,\Omega(x, \vec z_j +\vec y)^\dagger ~, \cr
\noalign{\kern 5pt}
&&\hskip -1cm
U_a' = \Omega(x, \vec y + \vec l_a)\, U_a\, 
   \Omega(x, \vec y)^\dagger ~, 
\label{newBC1}
\eeqn 
provided $\dd_M P_j' = \dd_M U_a' = 0 $.
The set $\{ P_j' \}$ can be different from the set $\{ P_j \}$.  
When the relations in (\ref{newBC1}) are satisfied, we write
$\{ P_j' \} \sim \{ P_j \}$~.
This relation is transitive, and therefore is an equivalence
relation.  Sets of boundary conditions form  equivalence classes of 
boundary  conditions.~\cite{YH2,HHHK,HHK}

\section{The Hosotani mechanism on orbifolds}

Wilson line phases are zero modes 
($x$- and $\vec y$-independent modes) 
of extra-dimensional components of gauge fields  
$A_{y_a} = \sum \onehalf A_{y_a}^\alpha
     \lambda^\alpha$  where   
$ \{ \lambda^\alpha , P_j \} = 0$  $(j= 0 \sim 3)$
 and $[A_{y_1} , A_{y_2} ] = 0$.
 The Hosotani mechanism concerning the dynamics of Wilson line phases
 are summarized as follows.~\cite{YH1,YH2,HHHK}
 
\begin{itemize}
\parskip=-2pt
\item[1.]
Wilson line phases  are physical degrees of freedom 
specifying  classical vacua.

\item[2.]
The effective potential $V_\eff$ for the Wilson line phases 
is nontrivial  at the quantum  level.  
The global minimum of $V_\eff$ determines the physical vacuum.

\item[3.]
If $V_\eff$ is minimized at nontrivial values, gauge
symmetry is spontaneously broken or enhanced.

\item[4.]
Gauge fields and adjoint Higgs fields (zero modes of $A_y$) are
unified.

\item[5.]
Higgs fields acquire finite masses at the one loop level.
Finiteness of the masses is guaranteed by the gauge invariance.

\item[6.]
Physics is the same within each equivalence class of boundary conditions.

\item[7.]
Physical symmetry of the theory is determined by matter content.

\end{itemize}

In this mechanism Higgs fields are identified with 
extra-dimensional components of gauge fields.  The expectation 
values of Higgs fields are determined dynamically.  It provides
dynamical gauge-Higgs unification.

\section{The $U(3)_S \times U(3)_W$ model of 
Antoniadis, Benakli and Quiros's}

The model~\cite{Antoniadis} is based on a product of two gauge groups 
$U(3)_S \times U(3)_W$ with gauge couplings $g_S$ and $g_W$ on 
$M^4 \times (T^2/Z_2)$.
$U(3)_S$ is ``strong'' $U(3)$ which decomposes to 
color $SU(3)_c$ and $U(1)_3$.  $U(3)_W$ is ``weak'' $U(3)$ which decomposes to  weak $SU(3)_W$ and $U(1)_2$.    
Boundary conditions  are given by 
$P_0 = P_1 = P_2 = I_S \otimes diag~(-1,-1,1)_W$,
which  breaks $SU(3)_W$ to
$SU(2)_L \times U(1)_1$ at the classical level.  
By the Green-Schwarz mechanism the symmetry reduces to 
$SU(3)_c \times SU(2)_L \times U(1)_Y$.  

\def\myb{{\vphantom{\myfrac{1}{2}}}}

The question to be answered is if the electroweak symmetry breaking 
occurs by the Hosotani mechanism. 
There are Wilson line phases in the $SU(3)_W$ group.
They  are 
\beeq
A_{y_1} = \pmatrix{&& \star\cr && \star\cr \star&\star\cr} 
=\pmatrix{~ & \myb \Phi_1 \cr
         ~~\Phi_1^\dagger ~ &  \cr}
         ~~,~~
 A_{y_2} 
=\pmatrix{~ & \myb \Phi_2 \cr
         ~~\Phi_2^\dagger ~ &  \cr} ~~.
\label{wilson4}
\eneq
$\Phi_1$ and $\Phi_2$ are $SU(2)_L$ Higgs doublets. 
The classical potential has flat directions.  Up to $SU(2)$ 
rotations they are represented by
$2 g_W R_1 \Phi_1^t = (0,a)$ and $2 g_W R_2 \Phi_2^t = (0,b)$.
The effective potential $V_\eff(a,b)$ is nontrivial at the 
quantum level.

With three generations of quarks and leptons $V_\eff(a,b)$ is 
evaluated to be

\begin{figure}[h]
%\begin{center}
\hskip .2cm
\includegraphics[width=6.5cm]{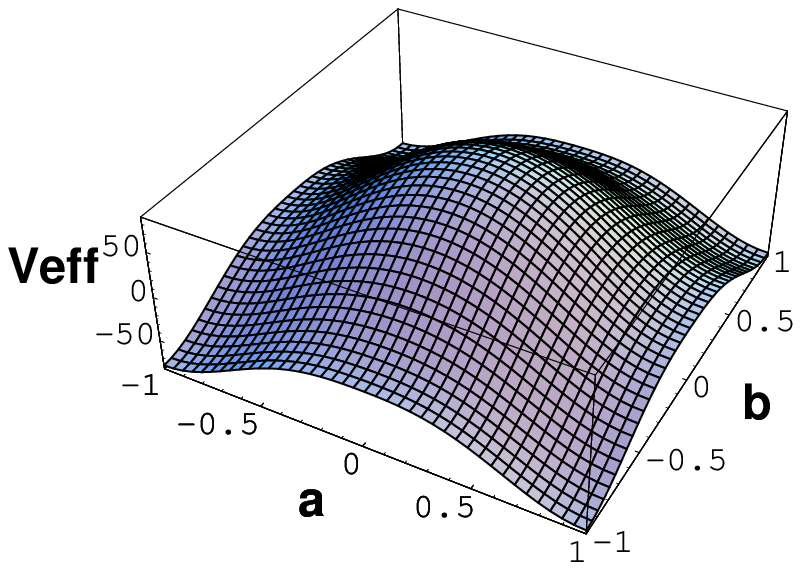}
\label{fig1}
%\end{center}
\vskip -4.0cm
\end{figure}

\beeq
\hskip 6.5cm 
V_\eff(a,b) =  - 40 \cdot I\Big( {a\over 2}, {b\over 2} \Big) 
    + 4 \cdot I(a,b)
\label{Veff1}
\eneq
\vskip 1.5cm

\noindent 
where
\beeq
I(a,b) = - {1\over 16\pi^2} \bigg\{
 \sum_{n=1}^\infty {\cos 2\pi n a \over n^6 R_1^6}
 +  \sum_{m=1}^\infty {\cos 2\pi m b \over m^6 R_2^6} 
 + \sum_{n=1}^\infty  \sum_{m=1}^\infty 
 {2 \cos 2\pi n a \cos 2\pi m b \over (n^2 R_1^2 +  m^2 R_2^2 )^3}
 \bigg\} ~~.
\label{Veff2}
\eneq
The global minimum of the effective potential (\ref{Veff1}) is located at
$(a,b) = (1,1)$, which corresponds to the $U(1)_{EM} \times
U(1)_Z$ symmetry.  Although the $SU(2)_L$ symmetry is partially
 broken and $W$ bosons acquire masses, $Z$ bosons remain massless.
 This result is not what we hope to obtain.  We would like to have
a model in which the global minimum of the effective potential
is located at non-integral values of $(a,b)$.  Some modification
is necessary. 

\section{Summary}
Dynamical gauge-Higgs unification 
is achieved in higher dimensional gauge theory.  Higgs fields are
identified with Wilson line phases in gauge theory.  Dynamical
symmetry breaking is induced by the Hosotani mechanism.
Finding a realistic model along this line is awaited.

%%%%%%%%%%%%%%%%%%%%%%%%%%%%%%

% A useful Journal macro
\def\jnl#1#2#3#4{{#1}{\bf #2} (#4) #3}

\def\Zphys{{\em Z.\ Phys.} }
\def\jssc{{\em J.\ Solid State Chem.\ }}
\def\jpsJ{{\em J.\ Phys.\ Soc.\ Japan }}
\def\ptps{{\em Prog.\ Theoret.\ Phys.\ Suppl.\ }}
\def\PTP{{\em Prog.\ Theoret.\ Phys.\  }}

\def\JMP{{\em J. Math.\ Phys.} }
\def\NPB{{\em Nucl.\ Phys.} B}
\def\NP{{\em Nucl.\ Phys.} }
\def\PLB{{\em Phys.\ Lett.} B}
\def\PL{{\em Phys.\ Lett.} }
\def\PRL{\em Phys.\ Rev.\ Lett. }
\def\PRB{{\em Phys.\ Rev.} B}
\def\PRD{{\em Phys.\ Rev.} D}
\def\PR{{\em Phys.\ Rev.} }
\def\PRe{{\em Phys.\ Rep.} }
\def\AP{{\em Ann.\ Phys.\ (N.Y.)} }
\def\RMP{{\em Rev.\ Mod.\ Phys.} }
\def\ZPC{{\em Z.\ Phys.} C}
\def\SCI{\em Science}
\def\CMP{\em Comm.\ Math.\ Phys. }
\def\MPLA{{\em Mod.\ Phys.\ Lett.} A}
\def\IJMPB{{\em Int.\ J.\ Mod.\ Phys.} B}
\def\cmp{{\em Com.\ Math.\ Phys.}}
\def\JPA{{\em J.\  Phys.} A}
\def\JPG{{\em J.\  Phys.} G}
\def\CQG{\em Class.\ Quant.\ Grav. }
\def\ATMP{{\em Adv.\ Theoret.\ Math.\ Phys.} }
\def\ibid{{\em ibid.} }

\vskip .8cm
\leftline{\small \bf References}

\renewenvironment{thebibliography}[1]
        {\begin{list}{[$\,$\arabic{enumi}$\,$]}  
% {\arabic{enumi}.}
        {\usecounter{enumi}\setlength{\parsep}{0pt}
         \setlength{\itemsep}{0pt}  \renewcommand{\baselinestretch}{1.1}
         \settowidth
        {\labelwidth}{#1 ~ ~}\sloppy}}{\end{list}}


\begin{thebibliography}{99}

\vskip -1cm 

\small

\bibitem{Manton1}
N.\ Manton, \jnl{\NPB}{158}{141}{1979}.

\bibitem{YH1}
Y.\ Hosotani, \jnl{\PLB}{126}{309}{1983}.

\bibitem{YH2}
Y.\ Hosotani, \jnl{\AP}{190}{233}{1989}.

\bibitem{orbifold1}
A.\ Pomarol and M.\ Quiros, \jnl{\PLB}{438}{255}{1998};

Y.~Kawamura, \jnl{\PTP}{103}{613}{2000}; \jnl{\ibid}{105}{999}{2001};

R.\ Barbieri, L.\ Hall and Y.\ Nomura,
   \jnl{\PRD}{66}{045025}{2002};
   \jnl{\NPB}{624}{63}{2002}.

\bibitem{Antoniadis}
I.\ Antoniadis, K.\ Benakli and M.\ Quiros,
     \jnl{\it New. J.\ Phys.}{3}{20}{2001}.

\bibitem{HHHK}
N.\ Haba, M.\ Harada, Y.\ Hosotani and Y.\ Kawamura, 
\jnl{\NPB}{657}{169}{2003};   
{\it Erratum}, {\it ibid.}  B{\bf 669} (2003) {381}.

\bibitem{YH4}
Y.\ Hosotani, hep-ph/0408012.

\bibitem{gaugeHiggs3}
N.\ Haba,  Y.\ Hosotani,  Y.\ Kawamura and T.\ Yamashita, 
\jnl{\PRD}{70}{015010}{2004}.

\bibitem{HNT1}
Y.\ Hosotani, S.\ Noda, and K.\ Takenaga, 
\jnl{\PRD}{69}{125014}{2004}.

\bibitem{HHK}
N.\ Haba,  Y.\ Hosotani and Y.\ Kawamura, 
\jnl{\PTP}{111}{265}{2004}.


%
%
\end{thebibliography}
\end{document}